\documentclass[namedreferences]{SolarPhysics}
\usepackage[optionalrh]{spr-sola-addons} 
\usepackage{graphicx}                    
\usepackage{color}                       
\usepackage{url}                         

\newcommand\FeI{\mbox{Fe\,{\sc i}}}

\newcommand{\aap}{    {\it Astron. Astrophys.}}

\newcommand{\apj}{    {\it Astrophys. J.}}
\newcommand{\apjl}{    {\it Astrophys. J. Lett.}}

\newcommand{\nat}{    {\it Nature}}

\newcommand{\solphys}{{\it Solar Phys.}}

\begin{document}

\begin{article}
\begin{opening}

    \title{Seeing-Induced Errors in Solar Doppler Velocity
    Measurements\footnote{The original publication is at \url{http://www.springerlink.com/content/n5v33354r06j2017/} }}

\author{Sreejith~\surname{Padinhatteeri}$^{1,2}$\sep
	R.~\surname{Sridharan}$^{3}$\sep
	K.~\surname{Sankarasubramanian}$^{1}$
	}
\runningauthor{Sreejith et al.}
\runningtitle{Seeing-induced errors in solar Doppler velocity measurements}

\institute{$^{1}$ ISRO Satellite Centre, Bangalore, India \\
			email: \url{sreejith.p@gmail.com}, \url{sankark@isac.gov.in} \\
           $^{2}$ University of Calicut, Kerala, India   \\
           $^{3}$ ESO, Chile
			email: \url{srengasw@eso.org} \\
	}
\begin{abstract}
Imaging systems based on a narrow-band tunable filter are used to obtain Doppler velocity maps
of solar features. These velocity maps are created by taking
the difference between the blue- and red-wing intensity images of a chosen
spectral line. This method has the inherent assumption that these two images are
obtained under identical conditions. With the dynamical nature of the solar
features as well as the Earth's atmosphere, systematic errors can be introduced
in such measurements. In this paper, a quantitative estimate of the errors introduced due to
variable seeing conditions for ground-based observations is simulated and compared with real
observational data for identifying their reliability. It is shown, under such conditions, that there is a strong cross-talk from
the total intensity to the velocity estimates. These spurious velocities are larger in
magnitude for the umbral regions compared to the penumbra or quiet-sun regions surrounding
the sunspots. The variable seeing can induce spurious velocities up to about 1 km s$^{-1}$.
It is also shown that adaptive optics, in general, helps in minimising this effect.

\end{abstract}
\end{opening}

\section{Introduction}
	\label{S-intro}

Recent observations have generated new interest in understanding the sunspot
fine structures like umbral dots, light bridges, and penumbral filaments
\cite{rimmele2008,rimmele2006a,scharmer2002,shussler2006}.
Detailed spatial and temporal observations of these fine structures are crucial
in understanding the physical mechanisms behind the formation of these structures.
With the success of the solar adaptive optics (AO), it is now feasible to
study structures close to the diffraction limit of modern
telescopes \cite{rimmele2004a,sankar2003,sankar2007,rimmele2008}.
These small-scale structures harbour flows that are important for understanding
the interaction between magnetic fields and plasma.
Observational study of these flows would provide constraints to the
theoretical models and to the magnetohydrodynamic simulations of these flows and
hence would lead to a better understanding of the overall structure.

Doppler shifts of spectral lines are regularly used to study the line of sight (LOS)
velocity of solar features. They are observed either with a spectrograph-based
instrument or with an instrument based on tunable narrow-band filter.
In instruments based on a tunable narrow-band filter, Doppler velocities are
obtained using the red- and blue-wing intensity images of a chosen spectral
line. Intensity at a fixed wavelength point in the blue and/or red wing of
any spectral line varies depending on the Doppler shift.
At a fixed wavelength point in the red wing of an
absorption line, a red-(blue-) shift will reduce (increase)
the intensity. Similarly, at a fixed point in the blue wing of an absorption
line, a red-(blue-) shift will
increase (decrease) the intensity.
Hence, the difference between the red- and
blue-wing intensities obtained at a fixed wavelength point is used to estimate
 the Doppler shift and hence the Doppler velocity. A magnetic field and its
gradients may affect the magnetically sensitive spectral line profiles and hence
the Doppler velocities estimated from them \cite{wachter2006,rajaguru2007}.
Therefore, magnetically insensitive lines
are preferred for a ``clean" velocity estimation. For example, \FeI\ 5434~\AA~or
\FeI\ 5576~\AA~are typically used to estimate Doppler velocities at the
photosphere. Systems based either on a tunable Fabry-P\'erot etalon or a universal
birefringent filter (UBF) (see \inlinecite{ubfreport1975},
\inlinecite{cavallini2006}, \inlinecite{stix2002}, and references therein for
details about these instruments) are used to obtain the required
spectral bandwidth (\textit{e.g.}, about 200m\AA~for photospheric spectral
lines). In both schemes, Doppler velocities are estimated from the difference between
intensities obtained at the blue and red wing of a chosen spectral line by
using the following relation:

\begin{equation}
    V = C.\frac{I_{\rm r} - I_{\rm b}}{I_{\rm r} + I_{\rm b}} ,
\label{eq1}
\end{equation}
where $I_{\rm r}$ and $I_{\rm b}$ are the red- and blue-wing intensities and $C$ is a
calibration constant which depends on the chosen spectral line and
the spectral resolution. $C$ can be obtained using a well-known procedure
\cite{Rimmele2004} briefly explained in Section~\ref{S-simul} of this paper.
With this definition, positive (negative) velocity correspond to flows towards
(away from) the observer. This sign convention is followed throughout this paper.

In such observations, the blue- and red-wing images are NOT
recorded simultaneously. The time difference between the two depends on the
wavelength tuning time, required number of wavelength positions, and the detector
read-out time. In most cases the detector read-out time, which is
typically a few seconds, limits the cadence. Hence, any appreciable
change in the observing conditions within this time interval can introduce
systematic errors in the velocity as well as spurious velocity structures.
If these spurious velocities and these structures are comparable to those of the
intrinsic velocities of photospheric structures, then the physical interpretation
of the structures will be ambiguous.
The typical intrinsic velocities in umbral dots are
of the order of a few hundred m s$^{-1}$, whereas the penumbral Evershed flows and
quiet-sun granular velocities are of the order of a few
thousand m s$^{-1}$ \cite{rimmele1995,lokesh2007}.

It is a well-known fact that ground-based observations are affected by
the atmospheric turbulence which is often characterised by Fried's
parameter ($r_0$) for long-exposure images. For this paper, a variable seeing
condition refers to the time variation of the parameter $r_0$. If the time
variation is a few cm within the few seconds required for obtaining
the red- and blue-wing images, then the variable seeing conditions can
introduce spurious velocity
signals. In ground-based observations, an adaptive optics system is used to minimise
the seeing effect. However, the performance of an adaptive optics system is a function of the
seeing conditions at the time of observations. \inlinecite{Rimmele2006} have shown that
the Sterhl ratio (one of the metrics for evaluating AO performance) of an AO
corrected image is a function of Fried's parameter ($r_0$).

\begin{figure}
\begin{center}
\includegraphics[scale=0.19,angle=90]{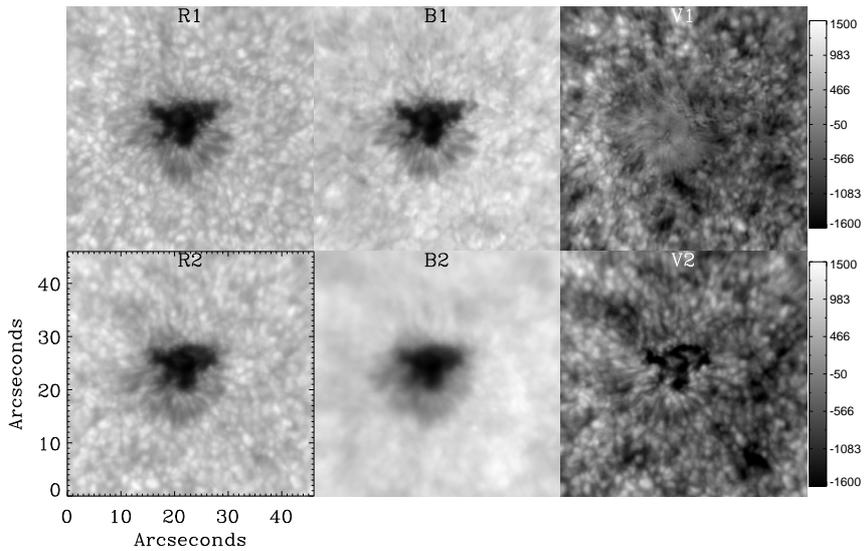}
\caption{Images showing the effect of variable seeing on velocity map. Top row images show
a sample good red- (R1), blue-wing intensity images (B1) and the velocity map (V1)
derived from these wing intensity images. Bottom row shows the same but observed during
variable seeing conditions.
}
\label{vel_image}
\end{center}
\end{figure}

An example of the effect of variable seeing is shown in Figure~\ref{vel_image}. The
observations were obtained at the Dunn Solar Telescope (DST) in Sunspot, NM,
USA using the UBF system. The images marked with R and B (R1, R2, B1, and B2) are
the red- and blue-wing images whereas the images marked with V (V1 and V2) are the
estimated velocities from the respective R and B. The top row images were obtained
under stable seeing conditions whereas the bottom row during variable seeing
conditions. The velocity map clearly shows spurious velocity signals in the
umbral and penumbral
regions under variable seeing conditions. This is also true with quiet solar
granulation (not shown here) but not so obvious to the eye due to the high intensity-velocity
correlation at these regions.

The aim of this paper is to quantitatively estimate the variable seeing-induced
spurious velocity signals, through simulations and to compare them with
observations. In Section~\ref{S-simul}, the method used to simulate the variable seeing
conditions is explained and the input data used for the simulation are also
explained. In Section~\ref{S-results}, results from the simulation using both space- and
ground-based data are discussed. The comparison of the simulation results
with the observed data and a summary are given in Section~\ref{S-summary}. We conclude
with the result that there is a good correlation between
seeing difference and spurious velocity signals, especially in the umbral
region of the spots. Our simulation indicates that spurious velocities can be as large as
1 km s$^{-1}$ and it is alarming to note that such high values are seen in the observed data.

\section{Simulations}
	\label{S-simul}

To simulate the effect of variable seeing on the Doppler velocities,
the red- and blue-wing images,
initially unaffected by the atmospheric turbulence, were convolved with
point spread functions (PSFs) produced using
different $r_0$. The PSFs were generated using the software tool Adaptive Optics Performance Evaluator (AOPE),
originally developed for performing simulations on the design needs of solar adaptive optics systems \cite{Sridharan2004}.

\subsection{AOPE}
	\label{SS-aope}
A detailed description of the effect of the atmospheric turbulence on the quality
of the images obtained with ground-based telescopes can be found in
\inlinecite{Roddier1981}. The instantaneous wavefront perturbations
induced by the atmosphere can be represented as a two-dimensional phase screen.
AOPE generates such phase screens following the Kolmogorov model of
turbulence, for any given value of Fried's parameter ($r_0$) and
derives a long-exposure PSF from them for a chosen
telescope diameter and observing wavelength. AOPE also simulates the
effect of the adaptive optics correction by fitting a model phase screen
with finite number of Zernike polynomials (which are generally used
in characterising the aberrations in optical systems) to the originally
generated phase screen and subtracting the best fit model phase screen
from the original phase screen.  The long-exposure PSFs
after adaptive optics correction are then generated from a series of residual
phase screens, again for a chosen telescope diameter and wavelength.
Thus, Fried's parameter, telescope diameter, number of equivalent
Zernike modes corrected by the adaptive optics system and the observing
wavelength are the input parameters to be selected by the users.
Long-exposure PSFs with and without a finite number
of Zernike-mode correction, and ideal PSF of the
telescope are the output parameters.  These output parameters
are characterised with the Strehl ratio, normalised Strehl resolution
and the strehl width which quantify the final image quality for a given
set of input parameters. In our simulations, we used  this tool to
obtain the ideal PSF of the telescope, and the PSFs with and without
the required Zernike correction. The corrected PSF depends on all the four input parameters,
whereas uncorrected PSFs do not depend on $Z$, the order of Zernike correction. The
ideal PSFs depend only on telescope diameter and wavelength $\lambda$.

\subsection{Input Data and Calibration}
		\label{SS-data}

The input data used for the simulation are obtained using the Solar Optical Telescope
(SOT) on-board \textit{\textit{Hinode}} - a satellite dedicated for solar observations. \textit{Hinode} is
a joint mission between the space agencies of Japan, United States, Europe, and
United Kingdom \cite{hinodemain}. Being a space-based instrument, the data
obtained from SOT are free from atmospheric turbulence. The narrow-band
filter imager (NFI) on SOT is used to observe wing
images of magnetically insensitive ($g=0$) \FeI\ line ($\lambda$=5576\AA~)
\cite{sotmain,sotref1}. The observations were carried out on
2007 July 14, 11:34 UT of an active region NOAA 10963. A pair of images
observed at the wings ($\pm$136m\AA~away from the line core) of the \FeI\
line 5576.09\AA~ are used as the input images for this simulation.

\begin{figure}
\begin{center}
\includegraphics[scale=0.32]{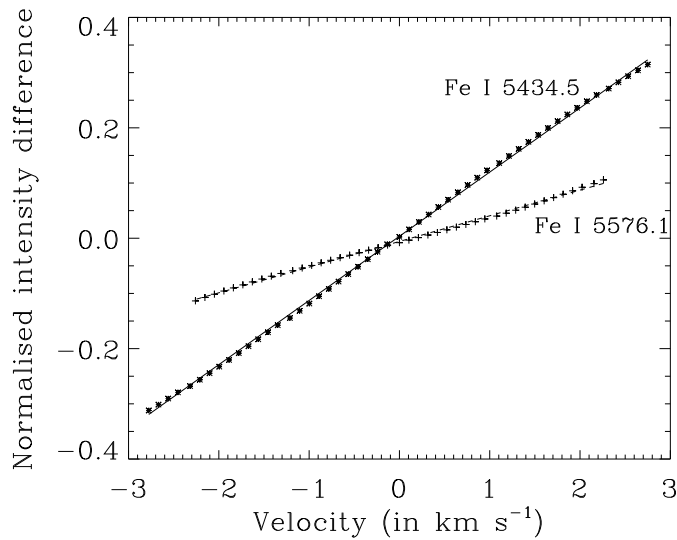}
\caption{Velocity versus normalised intensity difference curves estimated for the spectral lines
\FeI\ 5434 \AA~(wing points at $\pm$ 60 m\AA~; represented by asterisks)
and \FeI\ 5576 \AA~(wing points at $\pm$ 136 m\AA~; represented by pluses).
Solid and dashed lines are straight line fits to the respective curves.
}
\label{ubfcal}
\end{center}
\end{figure}

The input data for the simulation of ground-based images are obtained
using UBF observation of a sunspot carried out at the Dunn Solar Telescope (DST),
NSO, NM, USA, on 2005 December 28. The calibration constant $C$ ({\it cf.} Equation (1)),
required for deriving the velocity from the observed normalised intensity difference
$\delta I$ ( = $\frac{I_{\rm r} - I_{\rm b}}{I_{\rm r} + I_{\rm b}}$), is estimated using the spectral
profile from the Li\`ege atlas.
First, the atlas profile of the observed line is convolved with
a Gaussian filter profile of passband specific to the instrument used. In this
paper, data from the NFI on-board \textit{Hinode} as well as from the UBF at the Dunn
Solar Telescope are used. The passband is estimated to be 70 m{\AA} in the case of NFI and
142 m{\AA} for the UBF. The Doppler shift of the spectral line
$\Delta\lambda$ for a defined velocity $v$ is calculated by using
$\frac{\Delta\lambda}{\lambda} = \frac{v}{c} $
where $c$ is the speed of light. The convolved spectral line is then shifted
by an amount of $\Delta\lambda$ and the normalised intensity difference is calculated.
This is repeated
for a range of velocities. Figure~\ref{ubfcal} shows the relation between
$\Delta v$ and $\delta I$ for both 5576 \AA~(plus) and 5434 \AA~(asterisks) spectral lines.
The linear part of the curve is fitted with a straight line (solid and dashed lines
respectively) and 1/slope provides the calibration constant value $C$. For
large velocity values, either the line core or the continuum will cross one of the chosen
wing wavelengths and hence the $\delta I$ curve will deviate from the straight line.
This sets the limit for the velocity range that can be measured using this method. The
velocity range and the slope value (or calibration constant $C$) depends on the
spectral line and the chosen wing pair. This is clearly reflected in
Figure~\ref{ubfcal} in which the dynamic range achieved for \FeI\ 5576 \AA~
is smaller compared to that of \FeI\ 5434 \AA~.

\subsection{Procedure}
	\label{SS-procedure}
The simulation is carried out for a telescope diameter of 50 cm (commensurate
with \textit{Hinode}) and for different Fried's parameter values (starting from $r_0$ = 4 cm to
15 cm). In typical ground-based solar observations, Fried's parameter of 4 cm or
below is considered as bad seeing and an $r_0$ of 12 cm or above is
considered as an excellent seeing condition. For each $r_0$ values, PSFs with
and without Zernike corrections (of a particular order) are generated. For the
simulation, Zernike order is varied from 11 to 55 depicting different amount
of AO corrections. The PSFs generated are then convolved with the input image to
generate different observational conditions. In order to simulate the variable
seeing effects, the blue- and red-wing images are convolved with different
PSFs generated using different $r_0$ as well as different Zernike correction.

The velocities are derived using Equation (\ref{eq1}) for different simulated
conditions. The generated velocity images are used to quantify the effect
of variable seeing conditions. The umbra, penumbra, and quiet Sun regions
are analysed separately in order to quantify the effects in these different
regions.

For quantifying such variable seeing conditions between blue- and red-wing images,
we define the normalised seeing difference $(\delta r_0)$ by,

\begin{equation}
\delta r_0 = \frac{r_{\rm 0R} -r_{\rm 0B}}{r_{\rm 0R} +r_{\rm 0B}} ,
\label{eq2}
\end{equation}
where $r_{\rm 0R}$ and $r_{\rm 0B}$ are the seeing during the red- and blue-wing observations
respectively.

\section{Results}
	\label{S-results}

\begin{figure}
\begin{center}
\includegraphics[scale=0.25,angle=90]{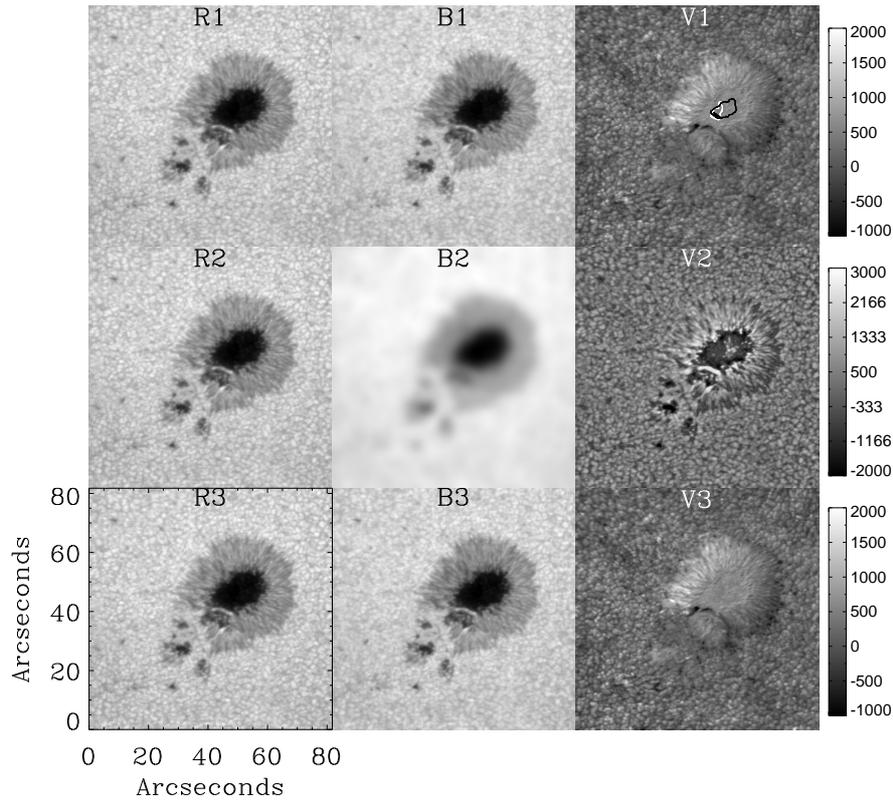}
\caption{\textit{Hinode} SOT-NFI images showing the effect of seeing difference on velocity.
This figure is similar to Figure~\ref{vel_image} except that it is simulated.
The top row shows the original red- (R1), blue-wing (B1) images and the corresponding
velocity map (V1). The middle row shows the original red- (R2), convolved blue-wing
(B2) images
and the corresponding velocity (V2). B2 is created by convolving original
blue-wing image  with a PSF of $r_0$ = 4 cm and without AO correction. In
the bottom row, B3 is created by convolving original blue-wing image with the PSF of
$r_0$ = 15 cm and $Z$=66. The colour bars on the right-hand side shows the grey
scale range for the corresponding velocity images. Dark and white contours in V1
are the regions selected for measuring $\sigma$ and $\mu$, respectively.
}

\label{hinode_fig2}
\end{center}
\end{figure}

The top row of Figure~\ref{hinode_fig2} shows a pair of red- and blue-wing images and
the velocity image used in our simulation.
The velocity image in the top row shows the
typical granular flows in the quiet region, Evershed flows in the penumbral region
and uniform or zero flows in the umbral region. The grey scale used for the
velocity images are also marked as a colour bar on the right side of the velocity
figures.
The bottom row shows the best case scenario when the seeing is good,
like with the case of $r_0$ = 15 cm and high order AO correction ($Z$ = 66). The estimated
velocity image looks similar to the original.
The middle row shows a seeing-affected blue-wing image simulated
for $r_0$ = 4 cm and without AO correction (the worst
case scenario), the original red-wing image (convolved with
ideal PSF), and the resulting velocity image. Note the change in the velocity
values and the appearance of velocity structures in the umbra. These small-scale
structures resemble umbral dots in the intensity image and hence are considered as
cross-talk from intensity.
This clearly shows the effect of the seeing difference inducing
spurious velocity structures as well as modifying the velocity amplitudes.
These effects are most notable in the umbral region.

\begin{figure}
\begin{center}
\includegraphics[scale=0.21]{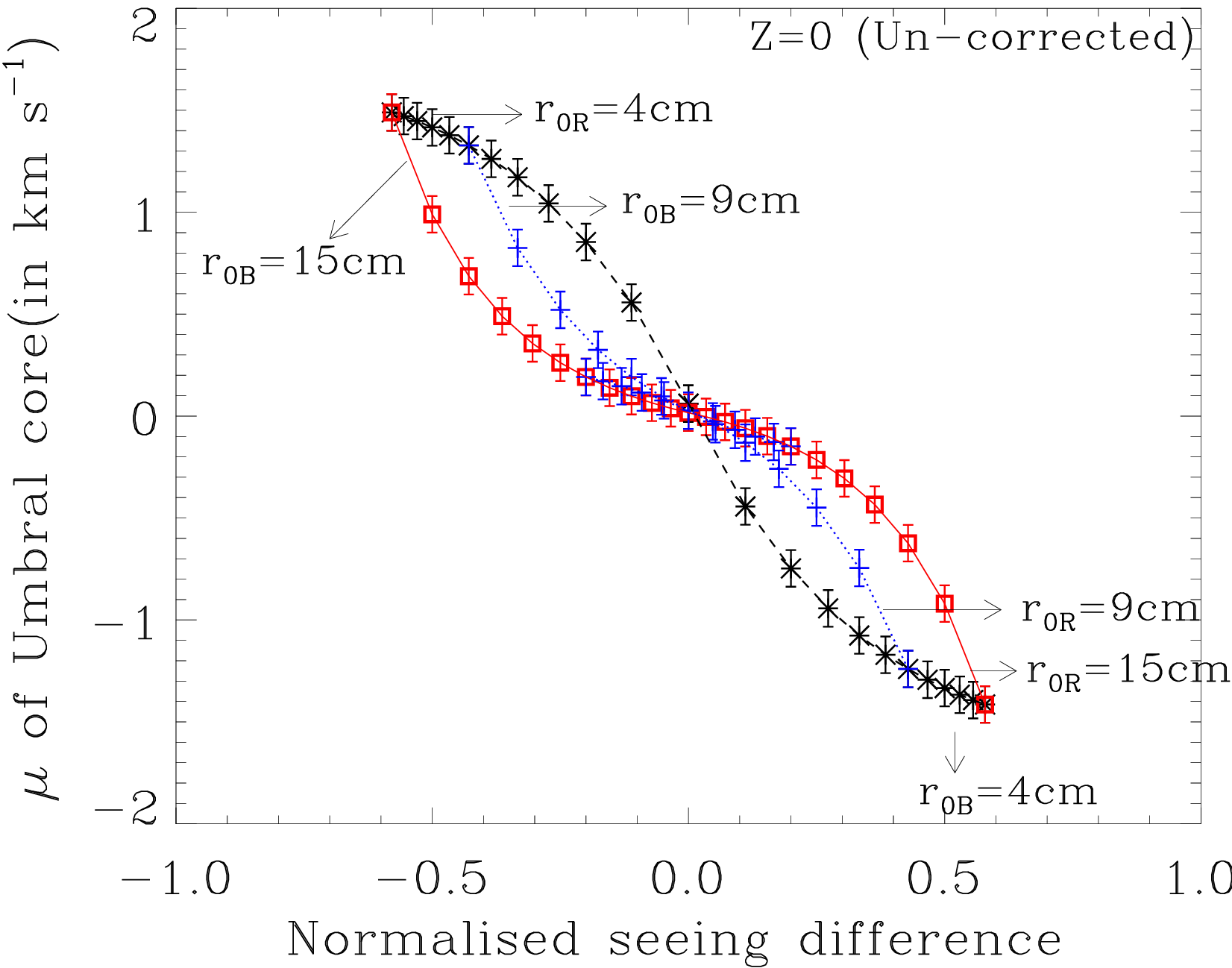}
\includegraphics[scale=0.21]{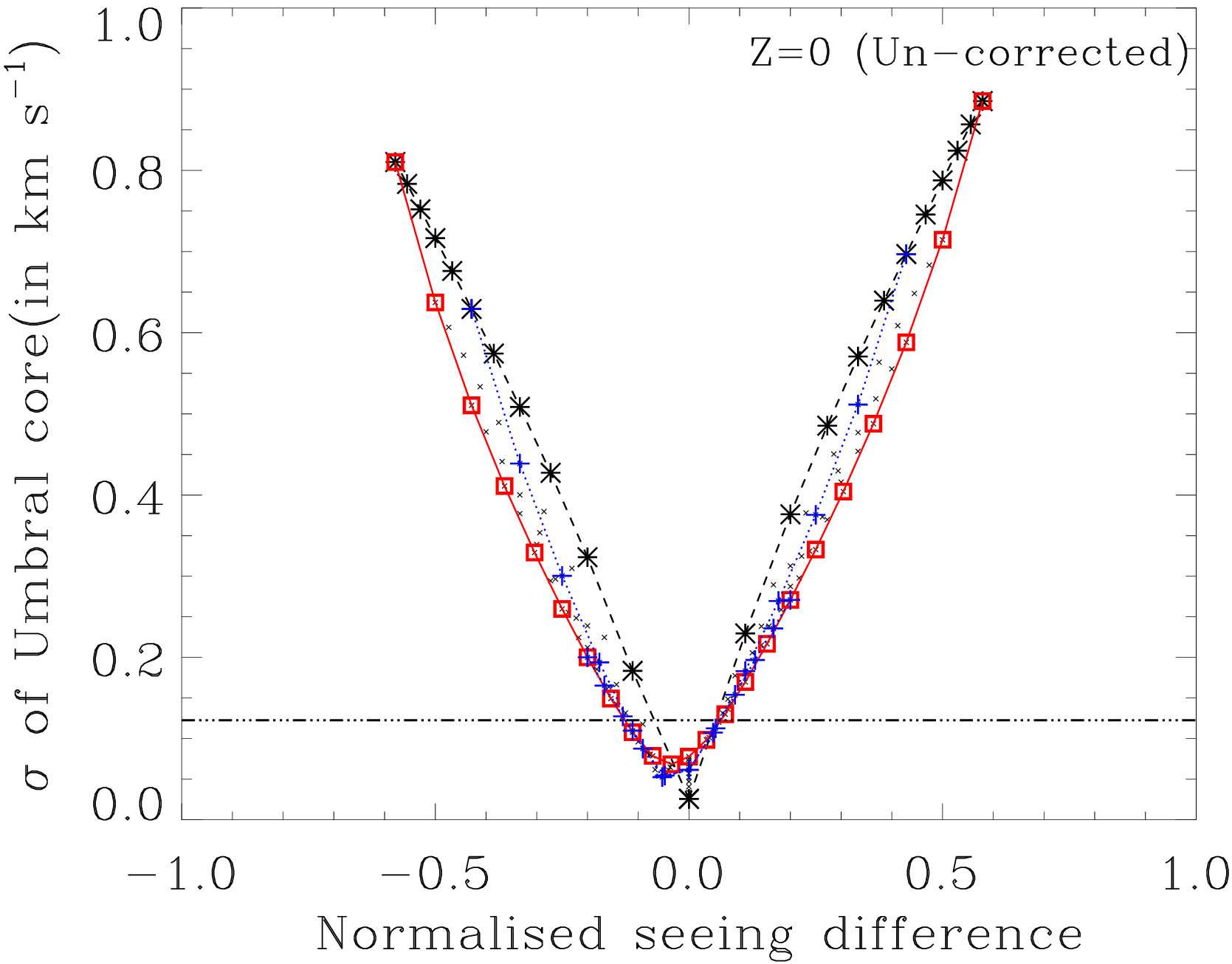}
\includegraphics[scale=0.21]{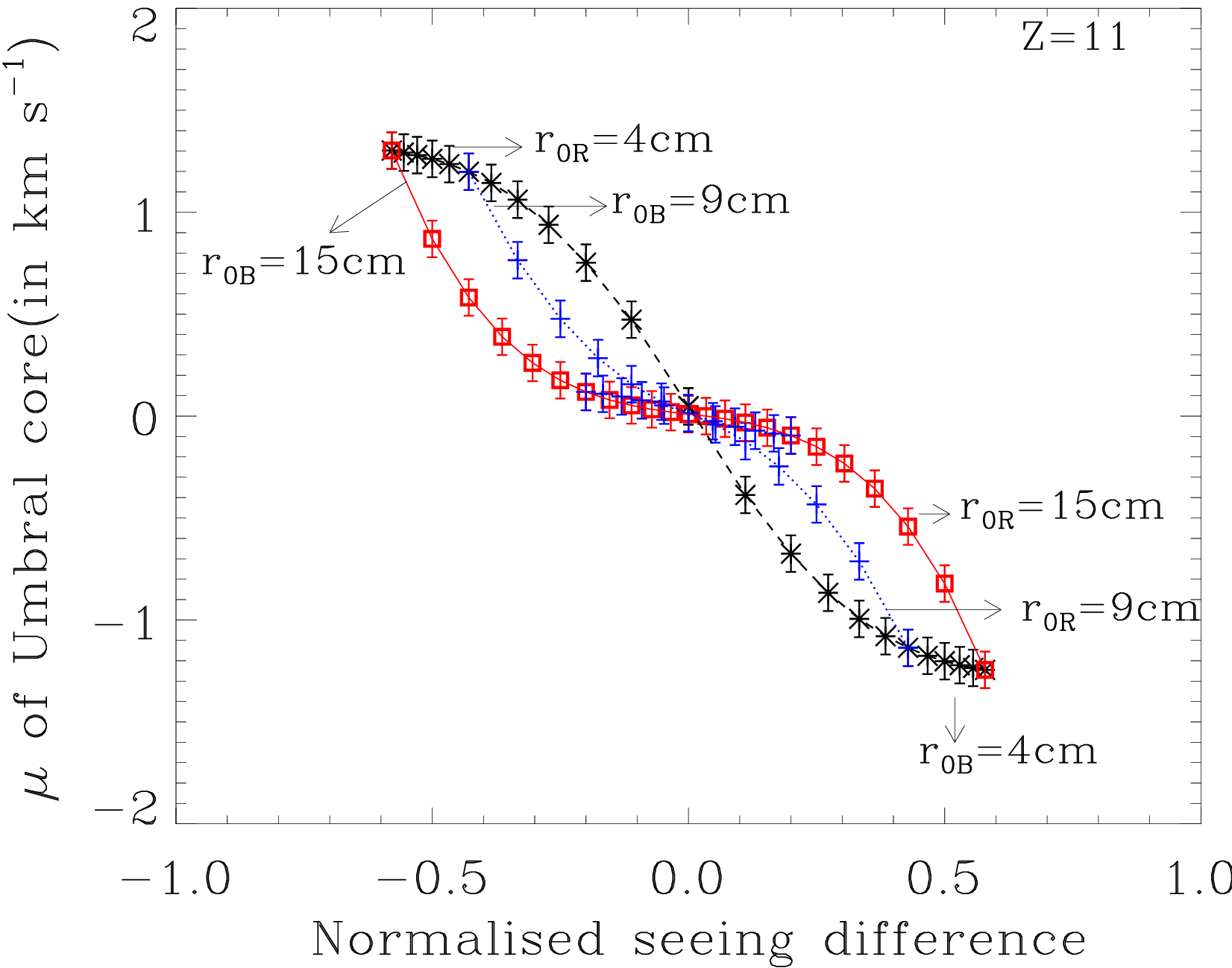}
\includegraphics[scale=0.21]{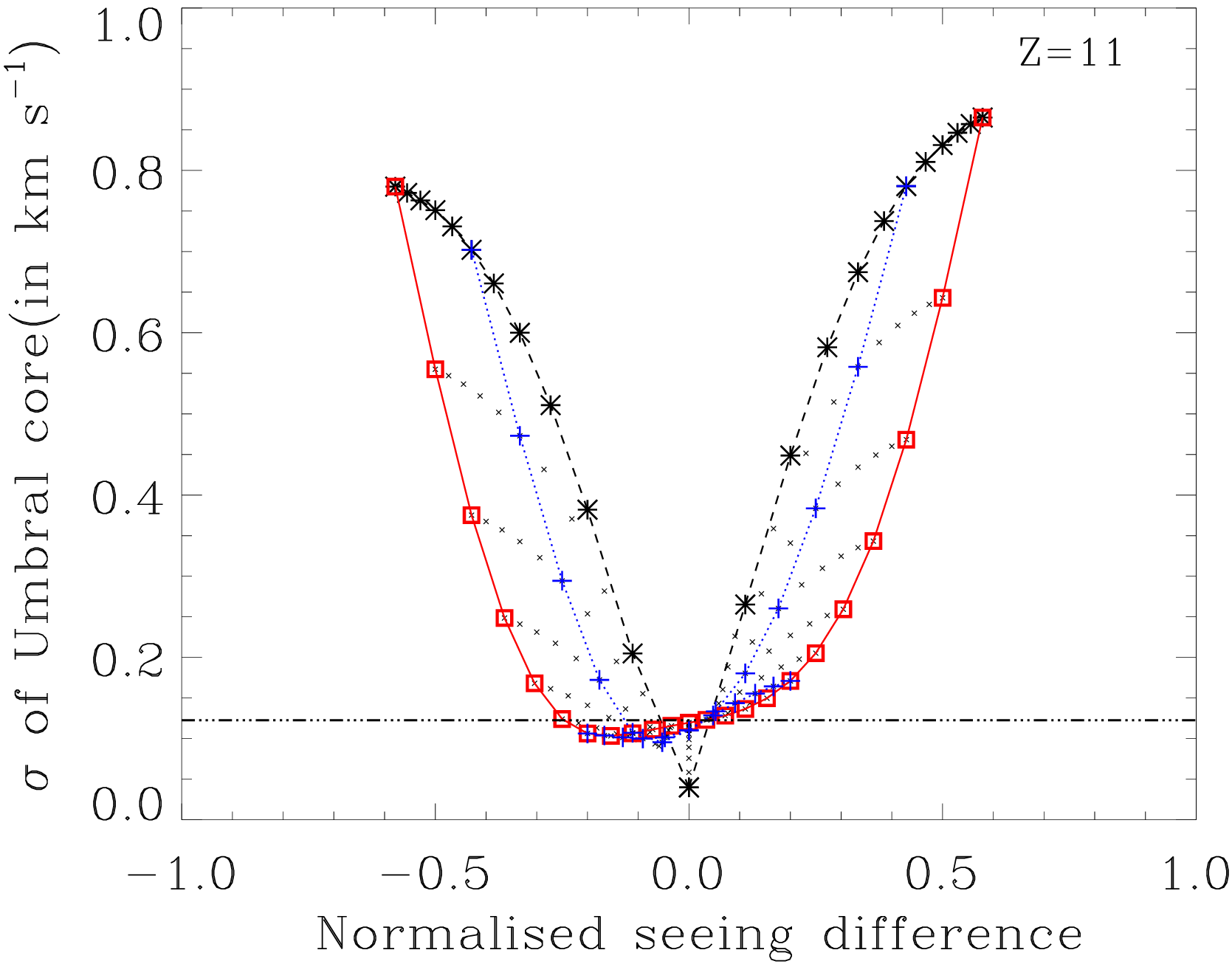}
\includegraphics[scale=0.21]{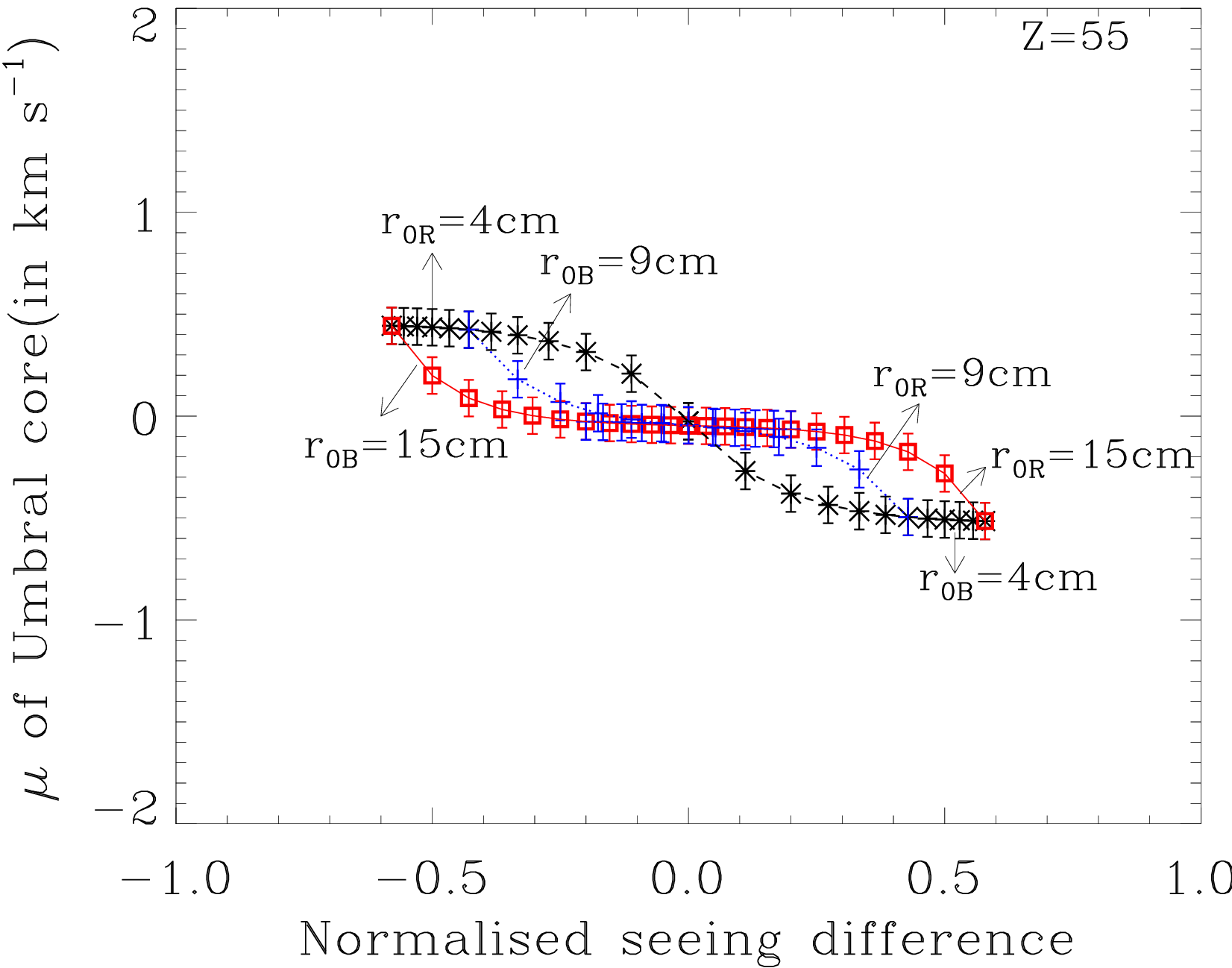}
\includegraphics[scale=0.21]{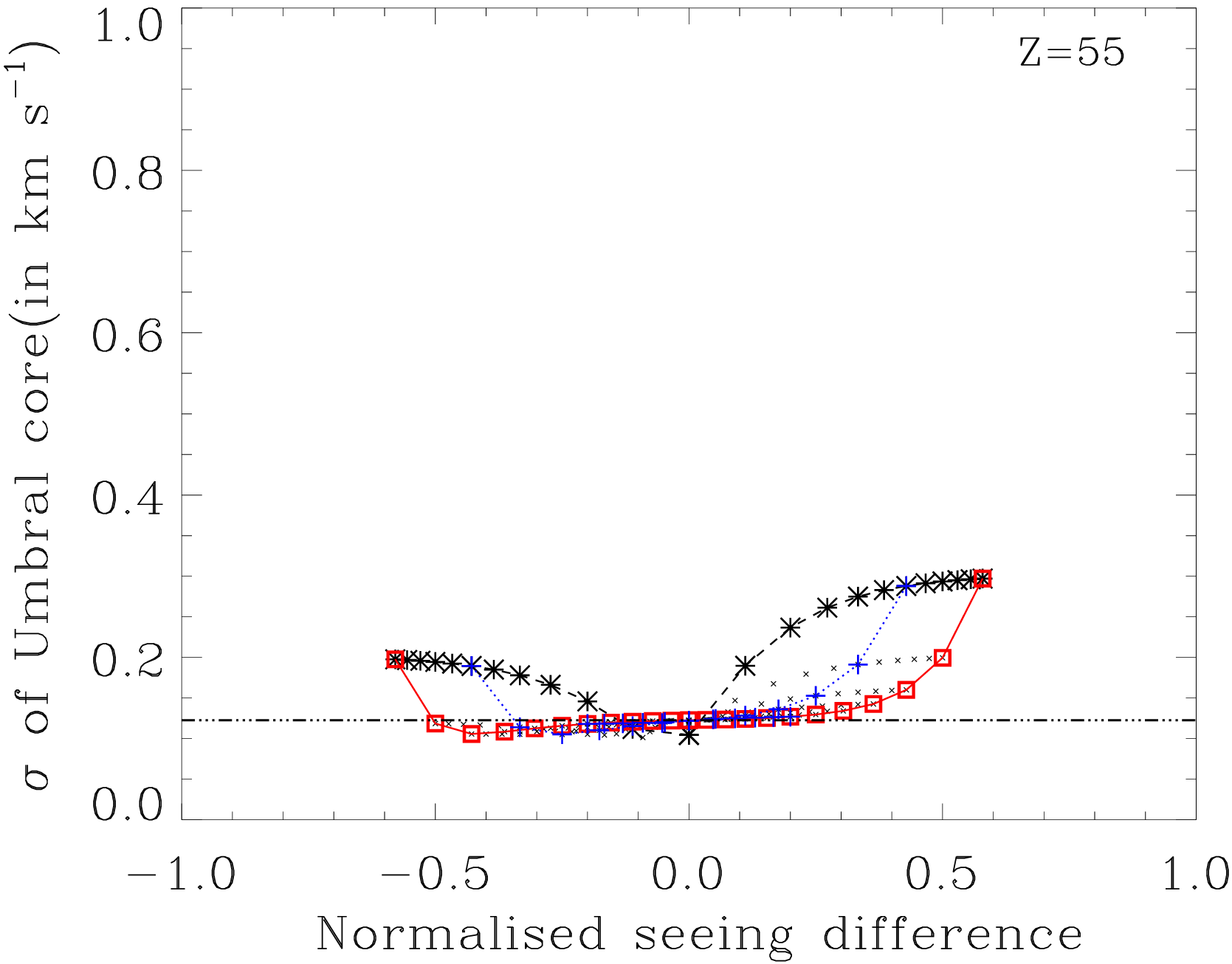}
\caption{ Effect of seeing difference on mean velocity (left column) and on velocity
structures (right column) in the umbral region. The top row is the case without AO correction,
the middle one with a Zernike correction of 11 and the bottom one with a correction of 55.
In all the plots, squares represent the best seeing case when $r_{\rm 0R}$
or $r_{\rm 0B}$ = 15 cm and the solid line connects the symbols, pluses are
for $r_{\rm 0R}$ or $r_{\rm 0B}$ = 9 cm, asterisks are for $r_{\rm 0R}$ or
$r_{\rm 0B}$ = 0 cm
and the dotted
lines and dashed lines connect the symbols, respectively.
}
\label{hinode1}
\end{center}
\end{figure}

In order to quantitatively study the effect on the mean velocity ($\mu$), a core umbral
region with minimal intensity structures (like umbral dots) is selected manually.
This region is shown as a white contour in V1 of Figure~\ref{hinode_fig2}.
The spurious velocity structures (as seen in V2 of Figure~\ref{hinode_fig2}) are
quantified using the standard deviation of the velocity of an umbral region which includes
the umbral structures (like umbral dots) and this is shown as a dark contour in
V1 of Figure~\ref{hinode_fig2}.
The two parameters ($\mu$ and $\sigma$) are plotted with normalised seeing difference
(cf. Equation (\ref{eq2})) in Figure~\ref{hinode1}. The seeing difference is
simulated by convolving one of the wing images (either red or blue)
with a PSF of a fixed Fried's parameter (called base seeing in this paper),
and the other wing is convolved with PSFs corresponding to variable Fried's
parameter ($r_0$ = 4 to 15 cm). This is repeated by varying the base seeing
values from 4 to 15 cm. The normalised seeing difference is calculated
as per Equation (\ref{eq2}), and the mean velocity is calculated at the selected core umbral
region. The same is also repeated with AO corrected PSFs, with Zernike correction order
varying from $Z$=11 to $Z$=55.

In Figure~\ref{hinode1}, the normalised seeing difference versus umbral core mean
velocity is plotted on
the left column. The top plot is the case without AO correction, the middle plot is with a
Zernike correction of $Z$=11 and the bottom one is for $Z$=55. All the
plots have three types of
symbols. Squares are for base seeing, where either $r_{\rm 0R}$ or $r_{\rm 0B} $ is equal to
15 cm and the other varying from 15 to 4 cm. In other words, one of the
wing images is affected by the best seeing condition and the other varies from best
to worst. Similarly, asterisks are for a base seeing of 4 cm and pluses are for 9 cm.
Each curve has two parts, separated by zero $\delta r_0$, and positive abscissa (
positive $\delta r_0$) means $r_{\rm 0R}~>~r_{\rm 0B}$. All the symbols are just connected
by lines of different line types.
As seen in the right column of Figure~\ref{hinode1}, the standard deviation ($\sigma$)
 varies with seeing difference and hence cannot be used as an error estimate for $\mu$.
The standard deviation of the velocity estimated in the
selected umbral region of the original velocity image is used as the error in the velocity
estimates for all simulated conditions.

All the three plots show that spurious velocity values increase with normalised
seeing difference.
For a particular normalised seeing difference value, there is a range of velocity
values depending on
the base seeing condition. A poor base seeing, or the dashed line in the plot
(connecting asterisks, where either $r_{\rm 0R}$ or $r_{\rm 0B}$ is equal to 4 cm),
represents the maximum limit, and a good base seeing, solid line
(connecting squares, where either $r_{\rm 0R}$ or $r_{\rm 0B} $ is equal to
15 cm) is the minimum limit of these induced velocities. Notice that the range
comes to almost zero at zero normalised seeing difference, irrespective of the
difference in the base seeing. As AO corrections are applied (middle and bottom plots),
both the maximum and minimum limit of the induced velocity decreases.

In the right column of Figure~\ref{hinode1}, the standard deviation of the selected
umbral area is plotted with normalised seeing difference. The top plot is without AO correction,
the middle one with a Zernike correction of $Z$=11 and the bottom one with $Z$=55. The horizontal line is the
value of $\sigma$ of the original velocity image. It is clear from the
curves that the seeing difference introduces spurious
velocity structures in the selected umbral region, and AO with
higher order correction helps in minimising these induced spurious
velocity structures. When both wing images are affected by similar bad seeing, the fine
scale intensity structures are smeared in both. Hence the intrinsic velocity contrast of
these structures is reduced representing the points below the horizontal line.

\begin{figure}
\begin{center}
\includegraphics[scale=0.20]{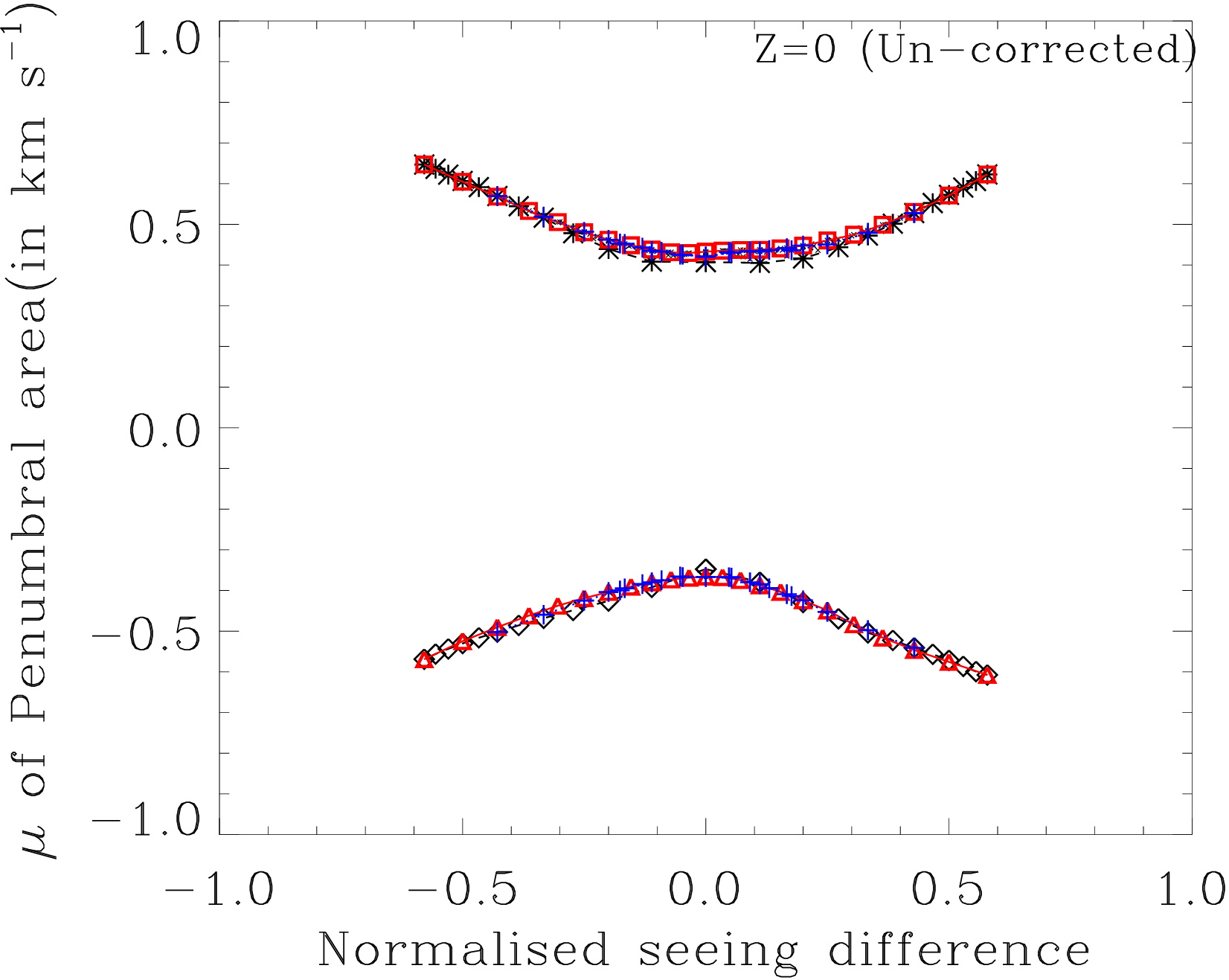}
\includegraphics[scale=0.20]{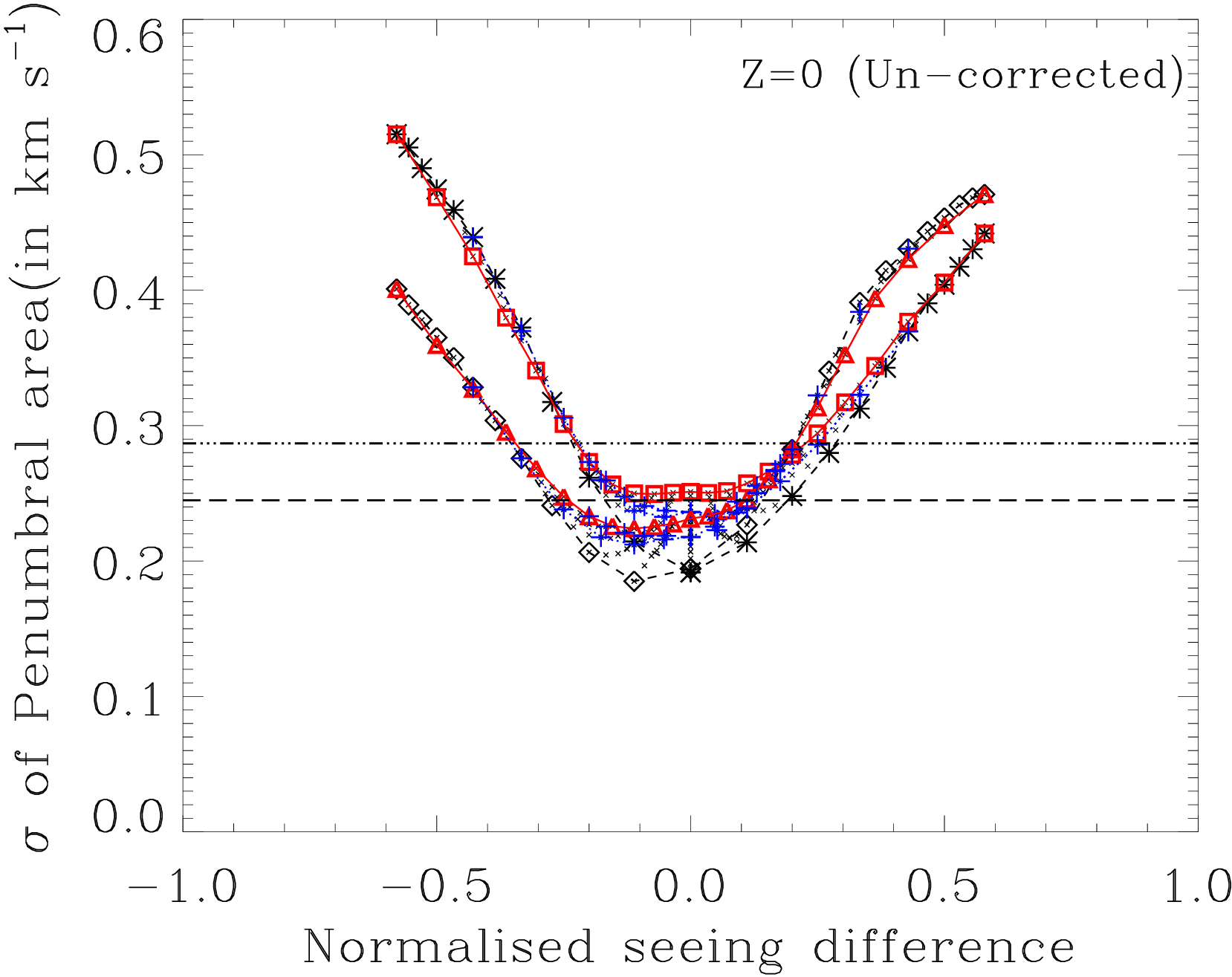}
\includegraphics[scale=0.20]{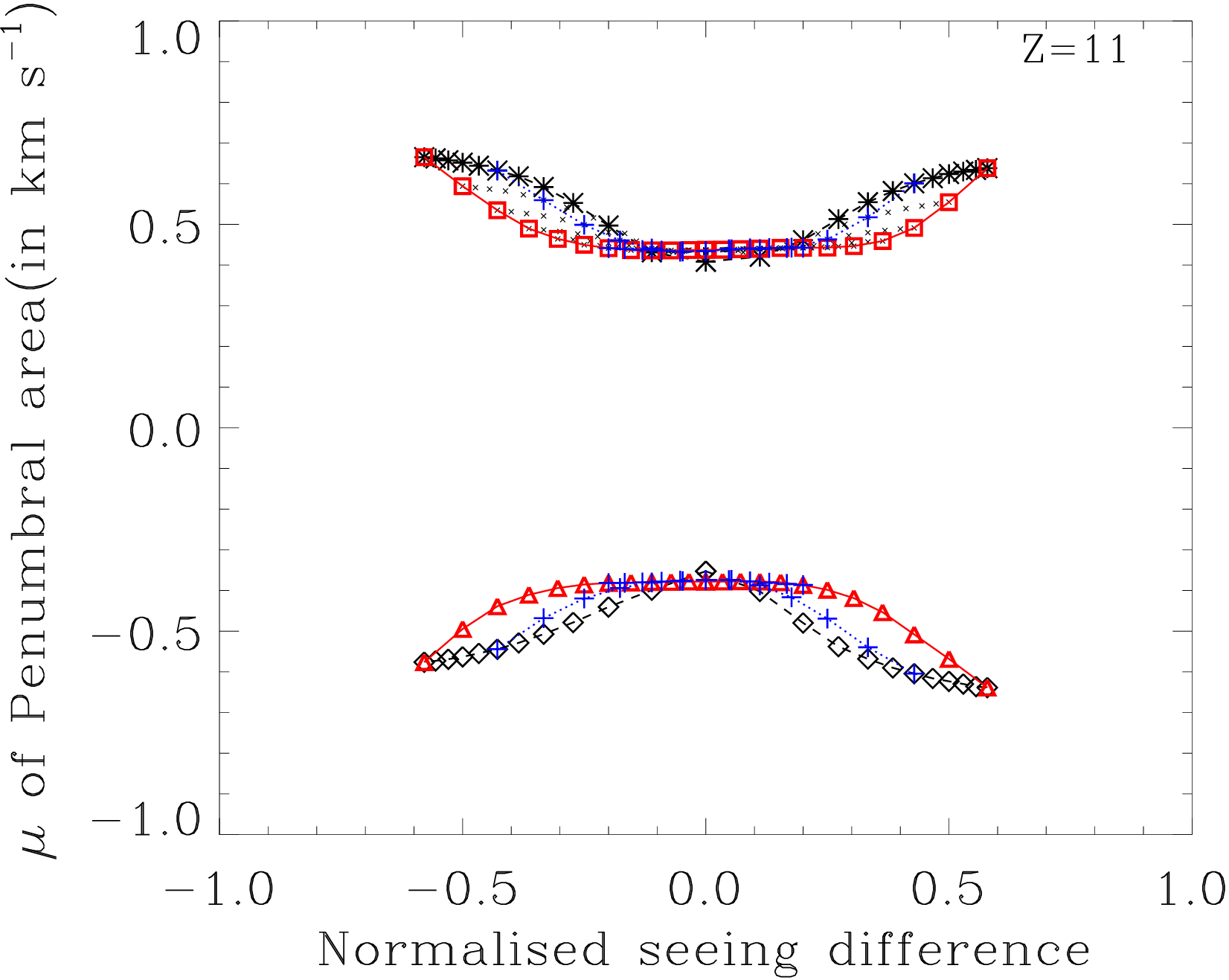}
\includegraphics[scale=0.20]{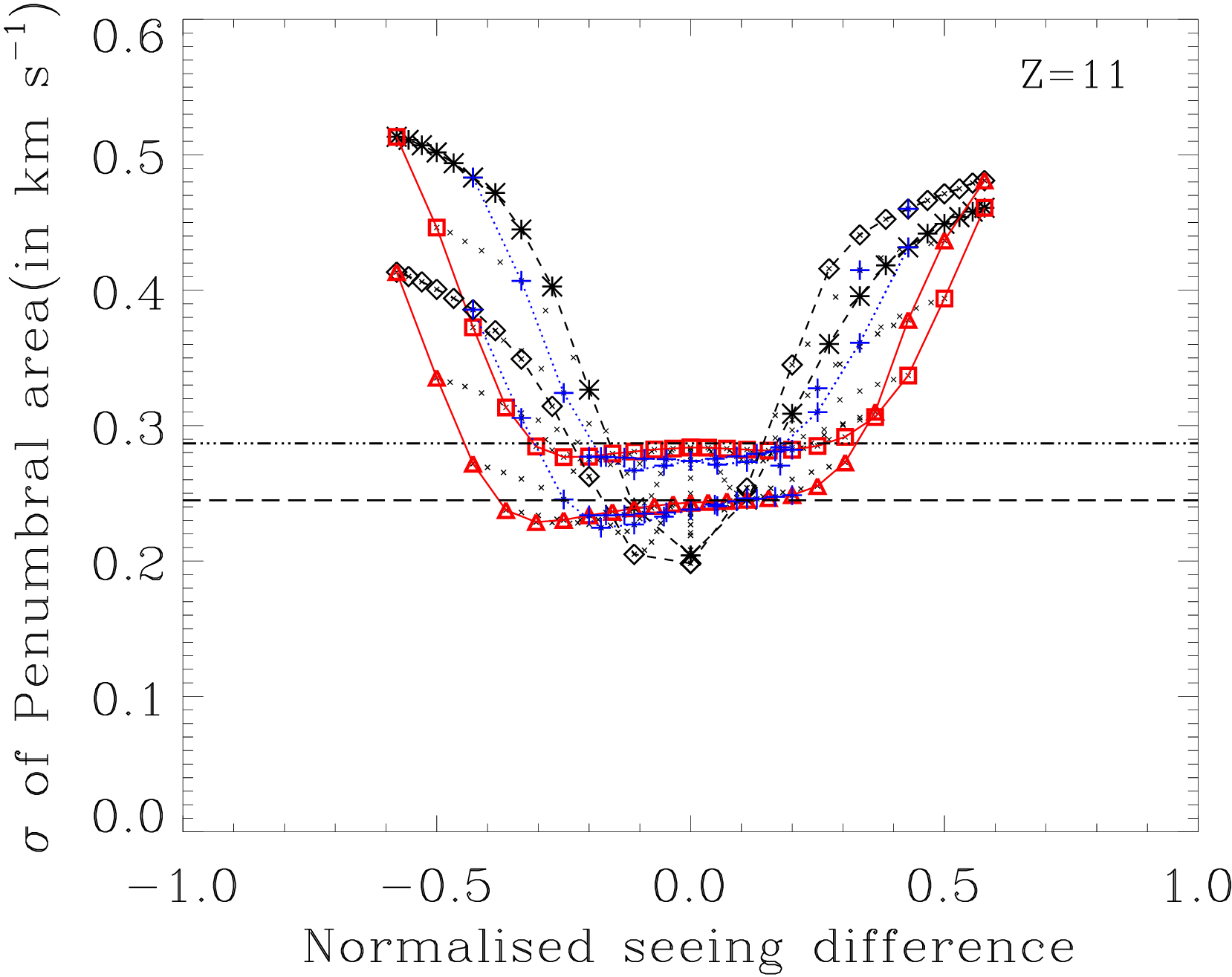}
\includegraphics[scale=0.20]{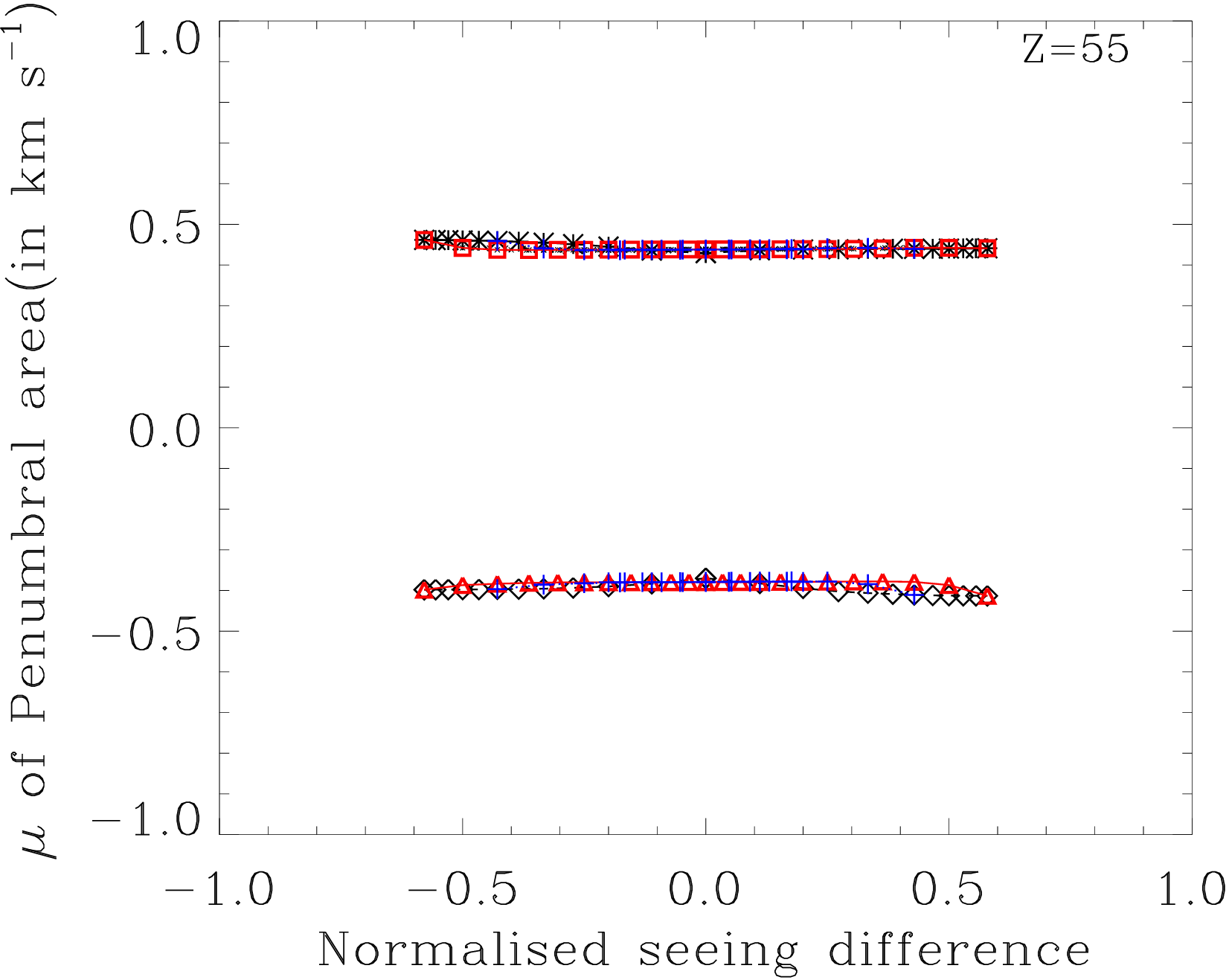}
\includegraphics[scale=0.20]{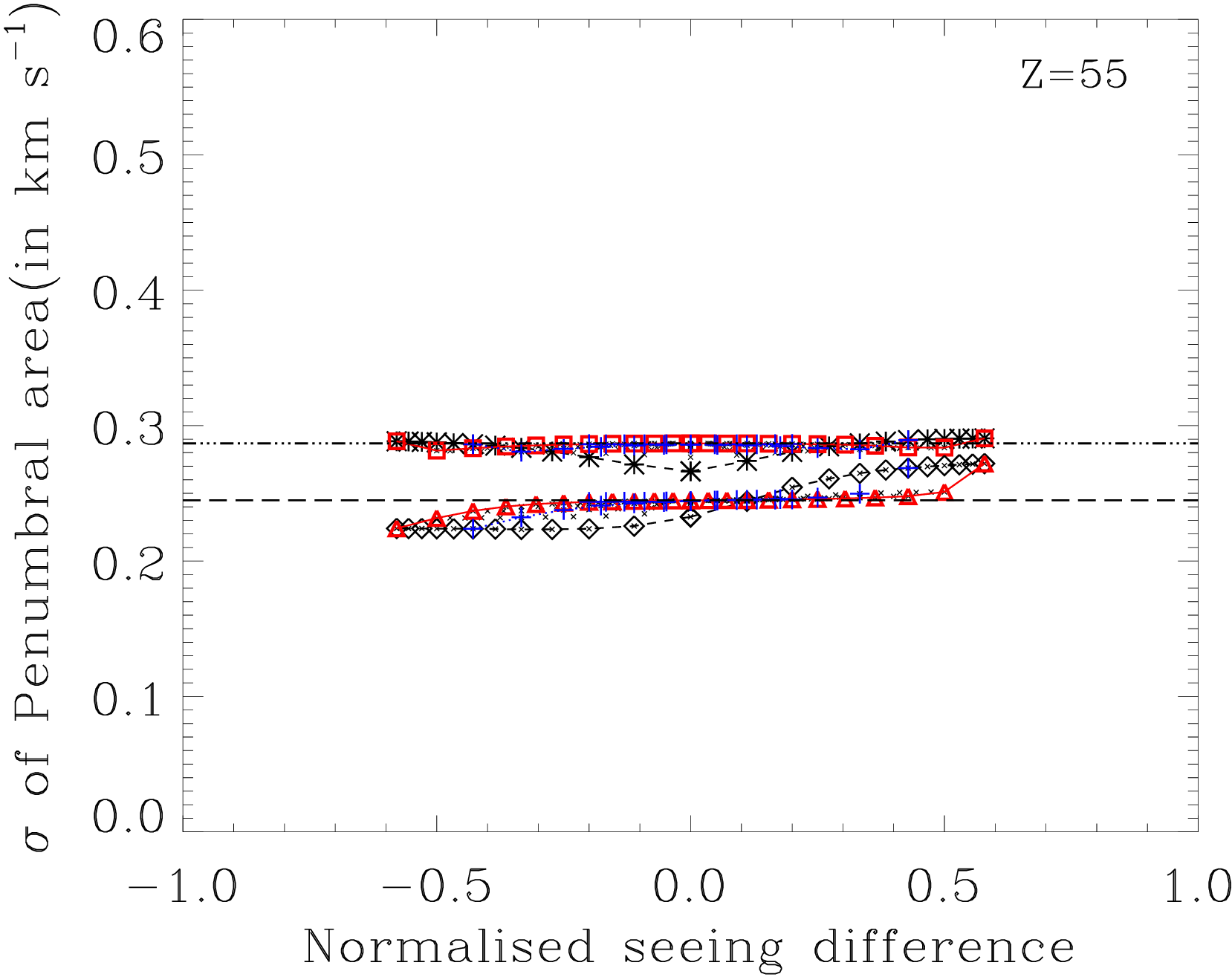}
\caption{Similar to Figure~\ref{hinode1} but for the penumbral region. The mean
velocity is calculated separately for positive (towards the observer) and
negative (away from the observer) cases.
Squares (positive side)  and triangles (negative side) represent the best base
seeing condition
($r_{\rm 0R}$ or $r_{\rm 0B}$ = 15 cm). Similarly, asterisks (positive side) and diamonds
(negative side) represent the bad base seeing condition (4 cm), and pluses
represent the medium seeing condition (9 cm). Lines are just a connection of data points.
}
\label{hinode2}
\end{center}
\end{figure}

A similar study is carried out for the penumbral regions. However, due to the
 presence of Evershed flows, the mean velocity ($\mu$) was calculated separately
for positive and negative velocities (flows away from and towards the observer).
The standard deviation of the selected penumbral area is a measure of penumbral structures.
These two parameters are plotted in Figure~\ref{hinode2} for different Zernike
corrections. It is clear from the figure that $\mu$ (left column) in a penumbra increases
very little (200-300 m s$^{-1}$) with normalised seeing difference,
whereas $\sigma$ (right column)
increases significantly. Similar to the umbral structures, with higher order AO
correction $\sigma$ falls off to original and also $\sigma$ does not fall to
zero due to the existence of penumbral velocity structures in the
input image. In the case of the penumbra, $\sigma$ of the input image cannot be taken
as the error in the velocity estimate due to the presence of Evershed flow velocity
structures, and hence the error bars are not displayed. Horizontal lines show
the value of $\sigma$ estimated from the initial velocity image.

The quiet sun-study showed a similar behaviour (with weaker amplitudes) like the penumbra and
hence is not discussed in this paper.

\subsection{Ground-Based Observations}
	\label{SS-ground}

A similar study is carried out with images obtained from ground-based observation.
The seeing was variable at the time of observation. About 3 s was
required to tune UBF from one wing to the other wing. Due to the variable seeing condition,
there were many wing-pair images affected by the seeing difference in the observed data
(one such an example is shown in Figure~\ref{vel_image}). From the observed data,
a pair of best images
were chosen as input for the simulations. Differential seeing was simulated as explained in the
previous section.

The curves of $\mu$ and $\sigma$ are very similar to the curves seen in
Figure~\ref{hinode1}, and hence they are not reproduced here in this paper.
The only notable difference was in the $\sigma$ curves and that can be attributed to the
difference in the umbral structures for these two different sunspot regions.

\section{Summary and Discussions}
	\label{S-summary}

In the velocity measurements from the ground using narrow-band filters,
there is always a chance of
systematic error in velocity estimates due to the seeing difference. If seeing changes within the
time of wavelength tuning from one wing to the other, it will create spurious velocity values.
We simulated such seeing difference conditions using the AOPE code, and we studied the induced
velocity values in different areas of a solar active region (like umbra, penumbra, and quiet
sun). It is concluded that the seeing difference affects the velocity estimates more in the umbra
than in the penumbral region of the sunspot. It is also seen that the effect is minimal in the
quiet sun. Under worst normalised seeing difference conditions, where one wing image is
obtained under
a good seeing condition ($r_0$ = 15 cm) and the other during worst condition ($r_0$ = 4 cm;
$\delta r_0$=0.58), the induced velocity is as high as 1 km s$^{-1}$. Such a large variation in seeing may
not occur often in real cases, but our sample observations (\textit{e.g.} Figure~\ref{vel_image}) using
the UBF at the DST have shown velocities up to 600 m s$^{-1}$.  It is also clear
that these induced velocity structures are due to cross-talk from intensity.
A correlation analysis carried out between
intensity and velocity shows a correlation coefficients up to 50\% in the worst case scenario.
These simulations also show that adaptive optics corrections can significantly
reduce such induced velocities, and velocity structures.

\begin{figure}
\begin{center}
\includegraphics[scale=0.3]{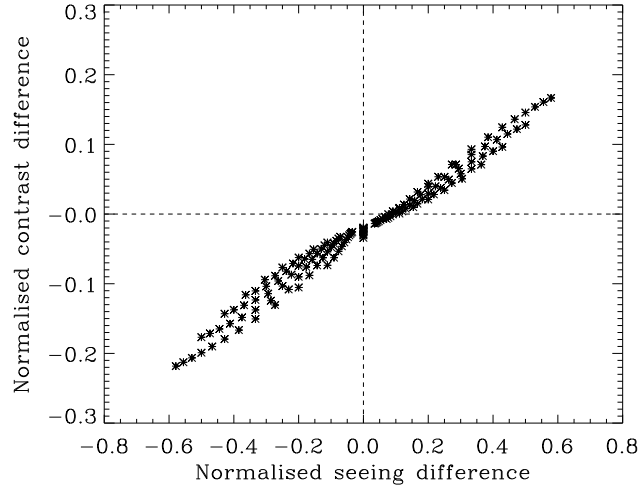}
\caption{ Scatter plot between normalised seeing difference and normalised
contrast difference  from the simulations using NSO data as the input.}
\label{dcontra}
\end{center}
\end{figure}

\begin{figure}
\begin{center}
\includegraphics[scale=0.3]{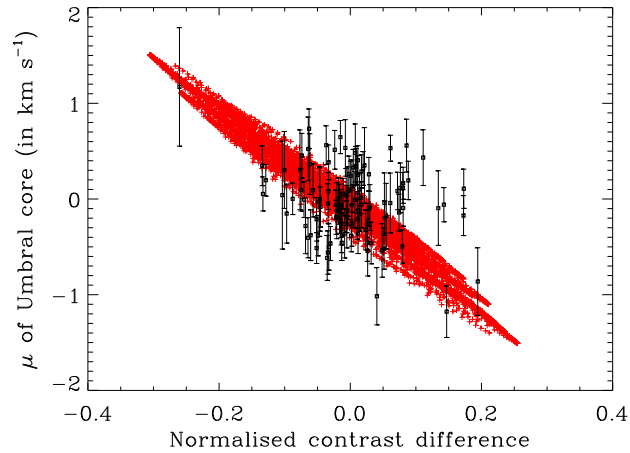}
\caption{Graph showing umbral core velocity with normalised contrast difference. The filled
squares are from the observation and red pluses are from the
simulation. The error bars on the observed points are three times the standard
deviation of the umbral core velocity distribution.}
\label{final_plot}
\end{center}
\end{figure}

We used UBF observations carried out during variable seeing conditions to compare with the
simulations. However, in order to compare with the real observations, contrast values are
used as a measure of seeing, since there were no simultaneous measurements of Fried's
parameter. The contrast at the umbra-penumbra border was used as a measure of seeing
rather than the quiet-sun contrast. This is because the lock point of AO was
in the umbra-penumbra border. The AO correction away from the lock point was really
variable due to the variable seeing condition and hence contrast obtained at
these areas will not represent the seeing changes alone in the umbra. We have also
taken care to choose the region which has a minimum velocity throughout the time
sequence of roughly 2~hours, as velocity can also change the intensity contrast.
We define contrast as the standard deviation of the selected region in the intensity
image. The normalised contrast difference is defined as the ratio between the difference
and total of the blue- and red-wing contrasts and hence is a proxy for the
normalised seeing difference value. Figure~\ref{dcontra} shows the relation
between normalised seeing difference and
normalised contrast difference from simulations with NSO data as the input.
It is clear from the plot that the spread in the curve will cause a degeneracy in the
normalised contrast difference  values for a particular normalised seeing difference.
Also, the zero normalised contrast difference does not fall on the zero
normalised seeing difference and vice versa.
This shift can be associated to the seeing difference in the input wing
images and to the convective blue-shift.

Figure~\ref{final_plot} shows the scatter plot between the normalised contrast difference versus $\mu$
at the core umbra estimated for the NSO data. Squares are from observation and
pluses are
from simulation. The simulation was extended to include all combinations of seeing values
varying from $r_0$ = 4 to 15 cm and a Zernike correction from $Z$ = 11 to 77 to mimic a real
observation. The error bar of the original data points are three times the
standard deviation ($\sigma$) of the velocity distribution in the selected umbral
core region. The value of this
$\sigma$ will be a combination of the actual error in the velocity estimation and the standard deviation
due to the induced structures because of the seeing difference.

In general most of the data points (70\%) fall within the simulated velocity distribution.
However, there are points well away from the simulated curve which may be due to
(1) the observed data being a time sequence spanning approximately 2 hours, and hence any evolution
can change both the velocity value and the contrast, (2) simulation assuming noiseless PSF after
perfect AO correction whereas in a real situation it will not be.

We conclude the folloeing. (1) A seeing difference introduces spurious velocity signals which are large in the
umbra compared to the penumbra and the quiet sun due to the low intrinsic velocities inside the
umbra. (2) The spurious velocity also depends on the base seeing condition.
If the general seeing conditions are very bad, then even a small seeing difference can
cause large spurious velocity. (3) The simulations have given a range of
spurious velocity,
for a particular seeing difference in very bad base seeing condition ($r_0$= 4 cm)
and for
very good base seeing condition ($r_0$ = 15 cm).  For a normalised seeing difference
condition of
0.5 this spurious velocity can range from 600 m s$^{-1}$ ( good base seeing) to 1200 m s$^{-1}$
(bad base seeing). A normalised seeing difference of 0.5 in the umbra  occurs when the ratio
of Fried's parameter between the red- and blue-wing images is larger than 3.
A normalised seeing difference of 0.5 corresponds to a normalised contrast difference
of approximately 0.2 and such values are seen in the observed data.
(4) With higher order AO corrections, the spurious velocities are
reduced by a factor greater than 4 for a normalised seeing difference of 0.5 or less for a
perfect AO system.

We wish to caution the reader that this simulation is just to approximate
the range of errors induced due to a seeing difference in velocity measurements
based on a narrow-band filter. An error in the range of 1 km s$^{-1}$ is important in areas like
the umbra and penumbra at the photospheric level. These results cannot be used to
remove the spurious velocities from the seeing-affected observed images.
At the same time, we cannot ignore all seeing-affected data,
especially in the case of observation of transient phenomena like penumbral formation
or a flare. In real observational situations, PSFs can vary even when $r_0$
remains constant (especially for exposures not long enough to average out the
seeing fluctuations).
In our simulation, it is assumed that the changes in PSFs are only due to changes
in $r_0$. If PSFs can be estimated using the AO data simultaneously along with the
Doppler measurements, then the variable seeing effects could be minimised by using
deconvolution techniques \cite{rimmele2006a}.
In the future, new methods and technologies may also be developed to observe the
blue- and red-wing images simultaneously to avoid such effects.
\acknowledgements{ \textit{Hinode} is a Japanese mission developed and launched by
ISAS/JAXA, collaborating with NAOJ as a domestic partner, NASA and STFC (UK) as
international partners. Scientific operation of the \textit{Hinode} mission is conducted by
the \textit{Hinode} science team organised at ISAS/JAXA. This team mainly consists of scientists
from institutes in the partner countries. Support for the post-launch operation
is provided by JAXA and NAOJ (Japan), STFC (U.K.), NASA, ESA, and NSC (Norway).
}


\IfFileExists{\jobname.bbl}{} {\typeout{}
\typeout{****************************************************}
\typeout{****************************************************}
\typeout{** Please run "bibtex \jobname" to obtain} \typeout{**
the bibliography and then re-run LaTeX} \typeout{** twice to fix
the references !}
\typeout{****************************************************}
\typeout{****************************************************}
\typeout{}}

\end{article}
\end{document}